# Jayant Vishnu Narlikar

*Naresh Dadhich*

It was late 1964; newspapers all over the country had a big frontpage splash, a young Indian don at Cambridge and his senior research collaborator had been able to see beyond Einstein in their new theory of gravitation. It had just been announced in the meeting of the Royal Society, London and the announcement had been enthusiastically received. This was precisely what the young nation was looking for as it was hungry for such recognitions and it was the first big one. For, an independent India was aspiring to catch up with the Western world in all spheres of life quickly, and more so in science, which is believed to be the key transformation vehicle for material progress. This is how Jayant Vishnu Narlikar (JVN) arose with a bang on the Indian science horizon. One does not have to stretch one's imagination much to know that he became a household name overnight, a science face of emerging and aspiring India, and an iconic role model. Exceptionally, he was awarded the civilian award, *Padma Bhushan* by the President of India. He is perhaps the youngest *Padma Bhushan* awardee.

JVN was born on 19 July 1938 at Kolhapur (then a princely state) in Maharashtra. His parents, Sumati and Vishnu Vasudev Narlikar (VVN) were both scholars, the former of Sanskrit and the latter, a legendary Professor of Mathematics at Banares Hindu University (BHU), Varanasi. JVN therefore had an ideal intellectual upbringing with Sanskrit providing a beautiful classic touch and flavour which most of us so sadly lack. It is therefore no surprise when he often effortlessly switches on to appropriate quotes from Sanskrit classics. VVN was one of the two persons, the other being N. R. Sen (Kolkata), who pioneered work in general relativity (GR) in the country. JVN thus had relativity in his 'janmaghunti' – the first taste of water at the birth, and he amazingly lived up well to remarkable initiation.

Right from school, JVN was an outstanding all-round student who excelled in all subjects as well as had equally engaging interest in sports. He played badminton very well. (Later he switched to tennis, and he and I used to play one set every morning in IUCAA, Pune, so long as we were both on the campus.) He grew up in Banares and had therefore imbibed a good bit of North Indian manners and language. This was combined with the Marathi directness and discipline inherited from his parents, which got further reinforced at Cambridge, not to mention that there was always a living Cambridge in VVN at home. The result of these cultural inputs made a wonderful plural mix of attitude and mannerism. The places of learning like BHU in the yesteryears fostered this wonderful social ethos. After completing B Sc at BHU, like his father JVN went to Cambridge as a Tata Fellow and excelled there too by finishing the formidable Maths tripos with flying colours in record time. At that time Fred Hoyle was the most sought after doctoral supervisor in astronomy and JVN's contemporaries included the most illustrious band in Stephen Hawking, Martin Rees, Brandon Carter and George Ellis, all of whom are among the best-known science faces of the day. From this extraordinary abundance of talent and brilliance, Hoyle chose JVN as his doctoral student, and this told a mighty lot on his reputation and standing as a student. It may also be mentioned that JVN won the distinguished Smith's Prize as a research student, and five years later, the prestigious Adams Prize in the august company of Roger Penrose and Stephen Hawking.

Like his mentor, Hoyle, JVN preferred to work in areas not fashionable but still fundamental. He continued to make valuable contributions to astronomy, although he deserved to receive greater appreciation for his work. As I will show in the following that he had half a dozen new ideas and predictions to his credit that were ahead of the times and hence were not taken seriously at that time. However, they have subsequently been accepted and verified observationally. In science the real measure of one's work is in creativity, in propounding a new idea or predicting some physical phenomenon and that gets subsequently accepted or verified. On this count his contributions certainly stand among the best, yet they have not attracted the attention they deserve.

There are two things here – observations may not always be sharp enough to give a decisive answer, but more importantly, all observations have to be interpreted within a framework of a theoretical model. It is the latter which is a very involved and complex issue. In all fairness and true to the spirit of truth-seeking, one should always keep one's mind open at this fundamental level. This stance, howsoever desirable and rational it may appear, is unfortunately sadly lacking in the present-day scientific community at large. One often sees all attention bestowed on bandwagon-type ideas to the exclusion of other viable alternatives.

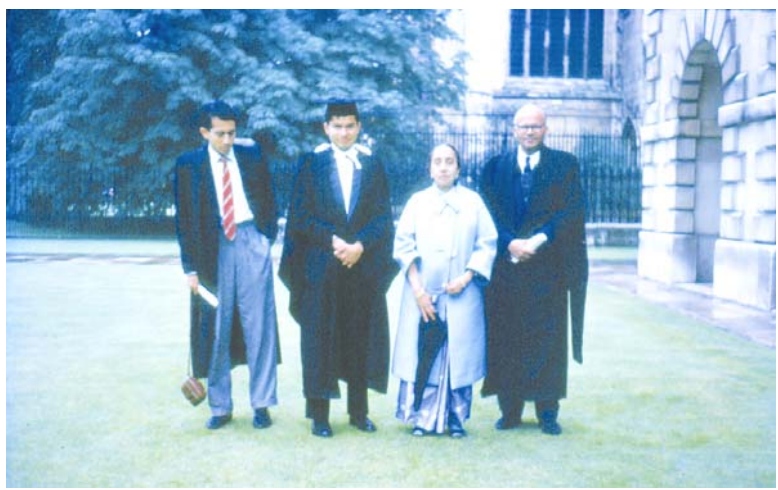

Jayant's graduation in Cambridge with his parents (Sumati and Vishnu Vasudev) and brother Anant.

# LIVING LEGENDS IN INDIAN SCIENCE

## Jayant Vishnu Narlikar

Jayant Narlikar was born on 19 July 1938 in Kolhapur, Maharashtra and received his early education in the campus of Banaras Hindu University (BHU), where his father Vishnu Vasudeva Narlikar was Professor and Head of the Mathematics Department. His mother Sumati Narlikar was a Sanskrit scholar. After a brilliant career in school and college, Jayant got his B Sc degree in 1957. He went to Cambridge for higher studies, becoming a Wrangler and Tyson Medallist in the Mathematical Tripos. He got his Cambridge degrees in mathematics: B A (1960), Ph D (1963), M A (1964) and Sc D (1976), but specialized in astronomy and astrophysics. He distinguished himself at Cambridge with the Smith's Prize in 1962 and the Adams Prize in 1967. He later stayed on at Cambridge till 1972, as Fellow of King's College (1963–72) and Founder Staff Member of the Institute of Theoretical Astronomy (1966–72). During this period he laid the foundations of his research work in cosmology and astrophysics in collaboration with his mentor Fred Hoyle.

Jayant returned to India to join the Tata Institute of Fundamental Research (1972–1989) where under his charge the Theoretical Astrophysics Group acquired international standing. In 1988 he was invited by the University Grants Commission as Founder Director to set up the proposed Inter-University Centre for Astronomy and Astrophysics (IUCAA). Under his direction IUCAA has acquired a world-wide reputation as a centre for excellence in teaching and research in astronomy and astrophysics. He retired from this position in 2003. He is now Emeritus Professor at IUCAA. In 2012 The World Academy of Sciences (TWAS) awarded him with a prize for setting up a centre for excellence in science.

In 1966, Jayant married Mangala Rajwade, a Ph D in Mathematics. They have three daughters, Geeta, Girija and Leelavati, all of whom have opted for careers in science.

Jayant is internationally known for his work in cosmology, in championing models alternative to the popularly believed big bang model. He was President of the Cosmology Commission of the International Astronomical Union from 1994 to 1997. His work has been on the frontiers of gravity and Mach's Principle, quantum cosmology and action at a distance physics. He has received several national and international awards and honorary doctorates. He is a Bhatnagar awardee, as well as recipient of the M.P. Birla Award, the Prix Janssen of the French Astronomical Society and an Associate of the Royal Astronomical Society of London. He is Fellow of the three national science academies as well as of TWAS. Apart from his scientific research, Jayant has been well known as a science communicator through his books, articles, and radio/TV programmes. For these efforts, he was honoured by the UNESCO in 1996 with the Kalinga Award.

Jayant recently broke new grounds in space research. Since 1999 he has been heading an international team in pioneering experiments designed to sample air for microorganisms in the atmosphere at heights of up to 41 km. Biological studies of the samples collected in 2001 and 2005 led to the findings of live cells and bacteria, thus opening out the intriguing possibility that the Earth is being bombarded by microorganisms some of which might have seeded life itself here.

Jayant was awarded *Padmabhushan* in 1965 (at the young age of 26), *Padmavibhushan* in 2004. In 2011, the Government of the State of Maharashtra gave him the State's highest civilian honour *Maharashtra Bhushan*.

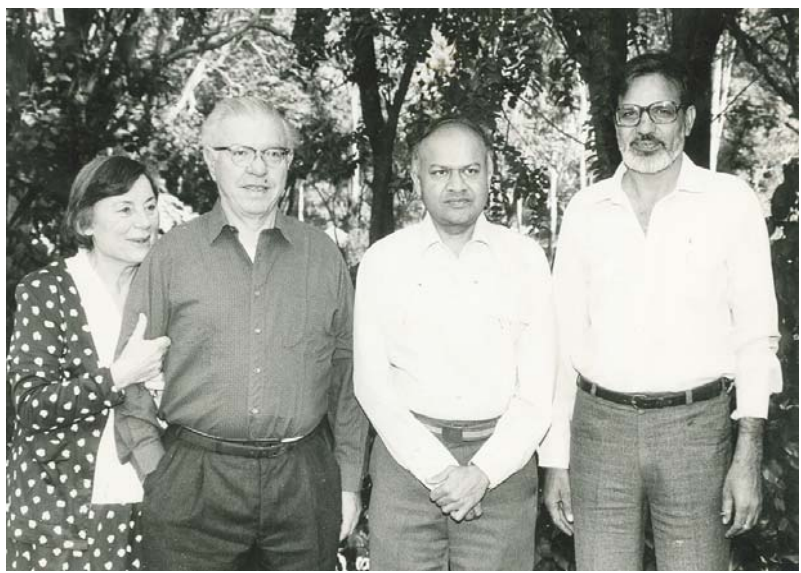

Jayant with Hoyles (Barbara and Fred) and the author.

One thing that stands out is that JVN has always been interested and worked on fundamental problems defying the strong peer group 'bandwagon syndrome' right from his graduate student days in early 1960s at Cambridge. As mentioned earlier, he probably acquired this trait from the legendary Hoyle, who was fiercely independent and enjoyed riding against the bandwagon. But more importantly, like Hoyle, JVN has a mind of his own and also the courage and conviction to challenge the established view if it does not, in his own assessment, stand the test of independent and dispassionate scrutiny. This is what he has done all through his scientific career. His subsequent professional isolation stems from this unflinching adherence to objective and dispassionate probing of facts and principle.

Though JVN is in a ridiculous minority (so much so he is now the lone member of the clan disbelieving the big bang theory) in following the path away from the mainstream beaten track, his critique of the standard established view of cosmology is taken seriously with respect and consideration. This is because his pioneering fundamental contributions could not be ignored and one cannot help acknowledging their originality and creativity. This is evidenced by his election as President of the Cosmology Commission of the International Astronomical Union for the term 1994–1997. This is a healthy trend and deemed necessary for growth of science in true spirit of truth-seeking. JVN is indeed a truth-seeker in the noblest classical tradition of scholarship. And in this role he stands out tall and alone.

Before I take up his scientific contributions, I must mention how JVN has pursued with great devotion and zeal a continuing discourse with people at large on scientific method and the enlightening and meaningful role science can play in their lives. His has been the most prominent science voice against astrology, superstitions and blind faith. He was the first and perhaps alone to come out strongly against UGC's proposal of introducing teaching of astrology in universities in the faculty of science. I should mention here that in his campaign he did have to put up with good bit of opposition and discomfort. His commitment to science communication and science education for public at large bears rich evidence in his popular science lectures at all levels and all through the country and abroad, interactions with school children in IUCAA's summer students programme and otherwise, and his popular science and science fiction books at large. He is certainly the man of Nehru's dream for creating a society with scientific temper. The other person who could share the stage of 'public scientist' with him is the ever affable Yash Pal, who has a charm and way of his own with people. I think we should all be aware of the debt we owe to people at large for their contribution to our comfortable living, and in return it is our bounden duty to participate in science communication in an effective manner so that people are better informed on matters of science and otherwise. This would help them in taking better informed and objective decisions and more importantly, to shed the burden of tradition-enforced blind faith and superstition.

Now I highlight some of JVN's seminal contributions to fundamental physics, astrophysics and cosmology. Let me open out with a pronouncement that many of his ideas were rather too radical for the time when they were propounded, but they have subsequently been proven right. Ironically even then, as mentioned earlier, they have often not received due appreciation and recognition for ingenuity, probity and priority. This is rather strange because usually one is lauded for anticipating and predicting forthcoming ideas and developments. Why has this not happened for JVN is something of a mystery.

Mach's principle (MP) for fundamental physicists is a fascinating enigma. Nobody fully understands and feels entirely comfortable with it; yet it has such an enticing appeal that none can ignore it. The real problem is in its statement which is ambiguous enough to present different faces to different adherents. There are as many versions of it as the number of people pecking on it. In its simplest form, it could be envisioned as saying that a particle exists because of existence of other particles in the Universe. It is envisaged that it is the mutual interaction between particles that is the cause for their collective existence. That is, a single particle cannot exist all alone. Going a step further, it is presumed that inertia of a particle arises from its interaction with the rest of the particles in the Universe. This is aesthetically attractive and has a great universal appeal. Even Einstein was greatly influenced by it and wanted to incorporate it in his theory of gravitation – general relativity (GR). He was much disappointed that GR did not do that as it admitted a solution for gravitational field of an isolated single particle.

In early 1960s, driven by the same inspiration as that of Einstein, both Hoyle and Narlikar (HN) wanted to construct a theory of gravitation which incorporated MP by writing inertia of a particle as an integral of its interaction with all other particles. Thus was born a new theory of gravitation – HN theory (HNT), which reduces to GR in the limit of many particles. However, the attention of the cosmological community got switched in the direction of the big bang cosmology by an important observation. In 1965, the isotropic cosmic microwave background radiation (CMBR) discovered, though accidentally by two Bell Lab engineers, Penzias and Wilson, unmistakenly pointed to the big bang cosmology. HNT was a brilliant theory and it was enthusiastically received by the theoretical physics community; unfortunately the observational interest went in the direction of the big bang theory which provided explanation of this observation. Before we go any further, it would be useful to understand in simple terms the phenomena of big bang and CMBR. Immediately after Einstein discovered GR, he applied his gravitational equation to the Universe as a whole for constructing a static model consisting of isolated particles like galaxies. Such a distribution is called dust, which is characterized by the absence of pressure. As there is no pressure to counteract gravitational attraction between the particles, such a Universe would collapse under its own gravity to a point singularity in no time. To counteract gravitational attraction, he then introduced the so-called cosmological constant which produced the required repulsive force to balance gravitational attraction. A few years later in 1924, a Russian physicist, Alexander Friedmann obtained a non-static solution for a homogeneous and isotropic Universe which was expanding or collapsing, and it did not need the cosmological constant for its existence. Independently, around the same time, it was also discovered by Lemaitre and later on generalized and refined by Robertson and Walker. This is the standard FLRW model of cosmology. After the discovery of non-static model of the Universe, Einstein lost interest in the cosmological constant. According to George Gamow, Einstein called this constant as the biggest blunder in his life. It is a different matter that about seven decades later the same cosmological constant is demanded by the supernova observations that point to accelerating expansion of the Universe. Among several exotic attempts to explain this cosmic acceleration, the most natural and satisfactory explanation comes from the repulsion produced by the cosmological constant.

The FLRW model is homogeneous and isotropic, which means the Universe appears the same from everywhere and in all directions; there is no preferred position and direction. As it expands different constituents in the form of galaxies and groups of galaxies–clusters are running away from each other. This means if we go back in time, they were closer, and in the limit they would all be riding on each other in a zero volume. At that event density of matter would be diverging to infinity and hence would be very hot, and that is what is the hot big bang beginning of the Universe. The Universe is born in a big bang explosion and it is expanding because of its initial impulsive momentum given by the bang. At the very beginning, it is believed that quantum fluctuations of vacuum produced matter in the form of fundamental particles like electrons, protons and photons–light, etc. It all comes out of nothing – vacuum. Initially the Universe is densely filled with these particles; light has very little free path; it keeps on bouncing from one to the other and getting absorbed and scattered. It cannot stream out of this cosmic soup of particles. The Universe is therefore perfectly opaque and dark. As it expands, it cools and then particles like electrons, protons and neutrons start combining to form atoms, that create free space for photons–light to stream out. This event, which marks decoupling of photons from other particles and is called the last scattering surface, is imprinted in the sky as a relic thermal radiation and its wavelength indicating the temperature. It cools as the Universe expands and thereby it loses energy, which means its wavelength goes on increasing; when it was caught by Penzias and Wilson, its wavelength was in microwave range, indicating temperature of 2.7 K (from absolute zero). This is what is the phenomenon of CMBR pervading the whole sky all the time in all directions. Its observation in 1965 by Penzias and Wilson was great advance in favour of the big bang theory, as it provided the strong observational support.

The first cosmology observation was that of Hubble's in 1929 of red shift of nearby galaxy, which indicated that it was running away. This was the observational support for the FLRW expanding model. Subsequently, with better telescopes available after the Second World War, red shifts of very large number of galaxies were observed, indicating that they were all running away from each other giving support to the FLRW expanding Universe model. Then, there was absolute lull in the observational front until the CMBR observation in 1965. This came in the backdrop of another competing cosmology theory – the steady state theory (SST) proposed by Hermann Bondi, Tommy Gold and Fred Hoyle in 1948. It was driven by the aesthetically beautiful and philosophically appealing perfect cosmological principle. It stated that the Universe always remains in the steady state as it offers the same view as observed from any location in space as well as at any time. The Universe therefore had no beginning and would have no end; it was eternally existing in the same state. Philosophically, it resonates wonderfully well with the Indian view of no beginninhg and no end, 'anaadi–anant', while it is in direct conflict with the Western view of creation of the Universe at some specific event in the finite past. On the other hand, big bang beginning as predicted by Einstein's GR resonates well with the Western conception of the Universe. It is interesting that the phrase 'big bang' was coined rather in jest by none other than Hoyle, the staunchest and the most formidable opponent of the conception.

SST was quite in conformity with the observations of those times and it then challenged observers to prove it wrong. It gave impetus for sharpening observational tools, particularly in the newly emerging science of radio astronomy which owed its remarkable progress in the following decades entirely to this challenge posed by the steady state theory. One of the main jobs of a theory is to challenge the existing view or framework, so that new tools are developed to prove it wrong or right. This is how science progresses.

The discovery of CMBR cast the die unambiguously in favour of the big bang and it was hard observational evidence which cannot be circumvented. It was difficult for SST to accommodate CMBR without giving up its most appealing perfect cosmological principle. Bondi was the first to acknowledge this fact. SST was a beautiful theory, if the observations did not support it, it had to go. Hoyle was not going to give up his favourite idea so easily; he and his former student Chandra Wickramasinghe argued most ingeniously to produce CMBR from iron whiskers in cosmic dust polarizing star light. It was a brilliant effort that could hardly convince his peers.

Besides CMBR, SST had a fundamental problem of maintaining uniform density of matter in an expanding Universe. For that, matter needed to be continually created so as to keep the steady state distribution. For big bang, it happened only once at the beginning, but here it had to happen all the time continually. HN introduced a creation scalar field with negative energy and stresses for creating new matter. The conservation of energy is satisfied by positive energy of matter created and negative energy of scalar field adding up to zero. Such an outlandish idea was rather too revolutionary in 1960s and completely out of tune with the times. Understandably, it was thought to be unphysical and thereby not taken seriously. Their viewpoint was that even at the big bang, matter came out of nothing violating the conventional conservation law, why could it then not happen continually all the time? By this way, the act of matter creation could, unlike an isolated big bang event which was not accessible to observations, be brought into the sphere of scientific enquiry and observation. This was a very valid standpoint. However, people were not convinced because they argued that the big bang was a special singular event of diverging energy density, while such special extreme conditions were hard to obtain everywhere all the time.

However, something similar is quite in vogue today, fashionably called 'phantom field' that is invoked for explaining the accelerated expansion of the Universe. Several of the phantom advocates do however acknowledge the origin of the idea in HN's creation field. Further, JVN also showed that creation field helped the initially inhomogeneous Universe to turn homogeneous and isotropic as it expanded. Two decades later, a similar result was christened as the 'cosmic no hair theorem' by Barrow and Stein-Schabes, who however did acknowledge the precedence of HN's creation field.

In 1966, HN proposed an ingenious cosmological model that radically departed from SST; it arose from a phase transition in the two states of the Universe, creative mode when there was creation of matter and non-creative mode when the creation field was switched-off. This happened in a bounded region of space which then expanded as a 'Friedmann bubble' in an external de Sitter Universe. This model anticipated the inflationary model by about a decade and a half, and it had been so acknowledged by the leading cosmologist Jim Peebles. It may be noted that currently inflation has been in hot news in relation to BICEP2 observation favouring one of the inflationary models. It predicted production of gravitational waves in the early Universe whose imprint was supposed to have been seen in the observed polarization of CMBR. The HN model did not receive the attention it deserved perhaps because of the unconventional nature of creation field. It is fair to say that equally bizarre phantom field has been invoked to explain an observed fact of accelerating expansion of the Universe. Similarly, the concept of inflation was invoked to establish causal connection between different parts of the Universe. This however does not take away their outlandish character, but they are however phenomenologically driven.

Yet another interesting application of creation field was to create matter in the centres of galaxies producing supermassive black holes of about a billion solar masses. They then controlled the size of galaxies, particularly elliptical ones. In 1966, JVN was the first to envision presence of supermassive black holes in the centres of the galaxies, which was soon afterwards also proposed by Donald Lynden-Bell and Martin Rees based on the conventional ideas and processes.

There is yet another pioneering work JVN did while he was a graduate student in 1961. Earlier, Gold and Hoyle had considered that matter was created in the form of neutrons which then decay producing electrons of high kinetic energy–temperature. Using thermal gradients so produced, they were able to create typical inhomogeneities in the Universe at the scale of 50 Mpc (1 Mpc is $10^6$ parsec and 1 parsec is approximately 3 light years). JVN worked on a model with this scale of inhomogeneities in the form of superclusters and voids. In this work, HN could successfully explain the observed distribution of radio sources as observed by the Cambridge radio astronomers then. Besides the analytical calculation, JVN had done numerical calculations on the then IBM 7090 computer which was perhaps the first attempt of calculation on a computer-generated Universe.

By far the intellectually most satisfying and exciting contribution of JVN is HN's generalization of the famous Wheeler–Feynman (WF) absorber theory of radiation proposed in 1945. WF developed the Maxwell electrodynamics as action at a distance theory, an alternative to conventional field theory. Maxwell theory, as is well known, is time symmetric and hence admits both advanced and retarded solutions. But we always observe retarded effect and never advanced effect. One has to break this symmetry and define an arrow of time. WF did it by invoking thermodynamics, the advanced component was absorbed by the other matter in the Universe and thereby defining an arrow of time.

It is interesting to note that interaction with rest of the Universe as envisaged in MP comes again in defining an arrow of time. The natural question then arose about asking this question in the framework of a gravitational theory that accommodated MP. This was what HN did and they argued that expanding Universe in cosmology provided a more natural arrow of time. For that first they had to generalize action at a distance formalism to curved space–time and then showed that cosmological expansion broke the advanced/retarded symmetry. It was remarkable that the correct result came out only for 'steady state and not for big bang' cosmology. Action at a distance formulation of Maxwell's electrodynamics in the cosmological setting clearly picked out steady state against big bang. This is indeed a remarkable result.

The next challenge was to quantize electrodynamics in this framework. Feynman attempted it but did not succeed and in the process discovered one of the most useful techniques of path integral formalism which had been so widely used. He thought that since quantization was not possible in action at a distance approach, field theory should be preferred over it. HN also tried their hand at the problem and were able to show that quantum considerations like spontaneous transition of atomic electron, pair production and annihilation, Compton scattering, etc. could be handled successfully and adequately in action at a distance framework. They further obtained the most remarkable profound result that response of the Universe constrained various integrals to give finite radiative corrections. Thus not only WF theory was quantizable, it required '*no renormalization*'. It was the back reaction of the Universe that provided a natural cut-off. This is a fascinating result by all counts, which has not attracted the deserved currency and appreciation, perhaps because it is based on the use of creation field and steady state cosmology.

One of the mysteries of quantum theory is renormalization which nobody understands, but it works. If there is a way to circumvent even if it requires invoking creation field and steady state cosmology, this is certainly a way forward and it should be so acknowledged. The famous WF paper was published in *Reviews of Modern Physics* in 1945. Half a century later in 1995, the journal invited HN to write a review article for telling the subsequent story of this very interesting and elegant formulation of electromagnetic interaction.

Further, HN also developed a similar approach to gravity and the resulting theory was connected to MP and reduced to GR under certain conditions. This is the HNT which I mentioned in the beginning. It also provided a background framework for the quasi-steady state theory which was to be developed later. In 1968, JVN also worked out a general correspondence under which a field theory could be replaced by a direct particle theory.

JVN was also among the first persons to apply quantum ideas to cosmology and had shown in 1977 that big bang singularity could be avoided by quantizing the conformal degree of freedom. It was quantization under special condition that quantum fluctuations of space–time were restricted to conformal degree alone. Then applying the Feynman path integral technique, he was able to compute exactly the probability of the Universe having a singular or non-singular beginning. It turned out that probability for the former was zero, while for the latter was unity. Thus the quantum gravitational effects provided a bounce to the Universe. This is perhaps the only calculation in quantum gravity which is exact without any approximation. In essence, the recent sophisticated treatments of this problem by application of loop quantum gravity have also been restricted by and large to the special case of quantizing conformal degree with some generalization to homogeneous Bianchi models. However, his pioneering step in this direction does not receive the attention and mention it deserves.

In response to the challenge thrown by the mainstream view – the party line, the famous rebellious trio – Hoyle, Geoffery Burbidge and Narlikar developed in 1993 the quasi-steady state theory of cosmology (QSSC) as their version of cosmology. One of the main issues was how to explain CMBR, which was by now a well-established fact. As mentioned earlier, Hoyle had argued that star light could get polarized by dust in the Universe to produce CMBR-like effect. For that dust should be there at all times. It was then envisaged that the Universe did not remain in a steady state at all times, but it went through cycles of very high dense state of high energy when matter got created, as it was done at the big bang. This was then followed by normal expanding phase, followed by collapsing phase to highly dense state. This is how it is in the quasi steady state. It is a cyclical Universe which goes through highly dense state resembling big bang and then it bounces back to expanding state. Matter from one cycle in the form of dust is carried over to the next cycle and that is how there is always dust required to polarize star light for creating a CMBR phenomenon. That is how they had proposed QSSC. It is creative and ingenious; yet several cosmologists are doubtful if the special dust required by the theory exists. Although experiments have conclusively shown that metallic vapours do condense into this kind of dust, the idea of its existence in space is considered incredible. It is ironical though that cosmologists have no qualms in accepting the presence of strange (non-baryonic) dark matter for which there is no laboratory evidence.

Steady state adhered to the most fascinating and beautiful conception in perfect cosmological principle which QSSC had to give up and thereby it lost all its aesthetic appeal. In QSSC, a big bang-like event is proposed continually and cyclically. It is now entirely driven by phenomenology and not by a compelling guiding principle. It was descending down from the pristine plane of concept and principle to the machine shop of phenomenology.

The picture that emerges is as follows: The Universe gets on the creation mode as alluded earlier during the dense phase and as it expands settles down to the normal steady state mode, creation gets switched-off. Then after some time it goes through collapsing phase to get to high energy creative mode again, and so the cycles go on indefinitely. In this process, there should also exist very old stars, which survived the collapsing phase, from the previous cycle and QSSC therefore predicts existence of stars older than the present age of the Universe, according to the big bang theory. This is a clear-cut prediction, but so far it remains unverified. If a star older than the Universe does show up, the big bang theory certainly goes; whether people would accept QSSC or not would be a different matter. It is important for a theory to make a clear prediction that could be subjected to observational test. On this primary count, QSSC stands perfectly well.

The trio had to work very hard to get their viewpoint across and the situation was appropriately summed up by the Editor of *Nature*, John Maddox in the editorial comment, entitled, 'The return of cosmological creation': '... they (the trio – Hoyle, Burbidge and Narlikar of QSSC) at least deserve credit for having pointed to one way in which the Big Bang, an event without a cause, might be brought within a wider framework'. This was an echo of Hoyle's words justifying the continual creation in the steady state theory about five decades ago. It was some recognition of the effort and ingenuity of the three distinguished rebels. With both Hoyle and Burbidge being no longer around, JVN who inherited it, remains the sole holder of the radical flag.

The final work of JVN concerns testing of hypothesis that the Earth may be bombarded by microorganisms from space. He proposed an experiment to be conducted by a joint effort of several laboratories. The experimental payload consisted of 16 stainless-steel tubes which could be opened and closed by ground command as they were carried up by a balloon. The tubes were initially evacuated and decontaminated and at specified heights ranging from 25 to 41 km, they had air pumped in using a cryopump. At different heights, different tubes were so filled. Then the payload was brought down and the contents of the tube were examined by two molecular biology laboratories. The latest study has been carried out by CCMB, Hyderabad and NCCS, Pune. Several bacterial species were found showing survival ability in UV, including three new species. Further studies are being planned to conclusively establish if these are from the Earth or from space. ISRO has been sponsoring these studies. It is purely an Indian experiment probing an interesting hypothesis.

Despite the great reputation JVN enjoyed globally among his peers, he was not given due credit for his several pioneering works which included the generalization of Wheeler–Feynmann theory, particularly back reaction of the Universe providing the natural cut-off, thereby eliminating the need for ever doubtable renormalization procedure, presence of supermassive black holes at the centres of galaxies, quantum cosmology and anticipating the idea of inflation by about 15 years, for ingenuity and priority. This is where sociology enters into the game – why do certain concepts or ideas even though new and radical gain currency and acceptance while some others do not? Perhaps the reason lies in the concept of creation field which is very difficult for conventional physics to swallow. True, now people have considered its cousin in phantom field with equally abhorring property of kinetic energy being negative, but they are driven to this desperate situation by observation. It is one of the ways amongst countably infinity that have been given the enticing name, 'dark energy' (anything we do not understand we term 'dark' – dark matter, dark energy and dark radiation) for explaining the observed acceleration of expansion of the Universe. It is a different matter that the most natural and satisfactory explanation is provided by the good old cosmological constant which had attained some kind of notoriety, though totally unjustified, as Einstein's greatest blunder. Though creation and phantom field are the same in spirit and substance, yet no observation except a theoretical model demanded the former, while the latter is invoked to explain an observation. I should however say that the result, back reaction of the Universe providing natural cut-off making renormalization unnecessary, should merit in favour of creation field and steady state cosmology almost as compelling as observation. It is this difference that is at the root of unfathomable aversion of people to creation field and the results based on it. Howsoever undesirable it may be, we have to accept this hard sociological fact.

JVN has been a brilliant global teacher through his marvellous books which are universally popular, and it feels great to hear people talking about his books with such admiration in all corners of the world. He has been a great mentor to people both younger and older than him. For instance, it was JVN who on his return to India in 1972, brought to light

A. K. Raychaudhuri (Presidency College, Kolkata), and the famous Raychaudhuri equation governing dynamics of Universe to Indian science community, even though this seminal work was done in 1955. He has built a vibrant school of theoretical astrophysics and cosmology and his students are world leaders in their respective fields. He is a scholar in the noblest truth-seeking tradition, that none of his students and colleagues who are mostly his students and friends work on his QSSC theory. He does not impose on others his ideas and instead encourages complete freedom and independence. This commitment to intellectual steadfastness and honesty is awe-inspiring.

The greatest gifts JVN has given to the nation and to science in general and astronomy in particular is IUCAA, a world-class astrophysics centre for promotion and growth of astrophysics teaching and research in universities. It was the most fascinating and rewarding experience for me to work with him in building up this wonderful institute right from its conception. He has an uncanny knack of making you share and be an equal participant in his dream and vision. There are few people who have this wonderful tact; the other person I have heard of was the founder-Director of IIT Kanpur, P. K. Kelkar, who would make even a security guard feel he was very valuable to the institute. This is the key to getting the best out of one's colleagues.

In mid-1980s, Govind Swarup conceived the Giant Metrewave Radio Telescope and was looking for an appropriate site for the same. I helped him with the logistics for site survey around Pune. In view of such major observational facility, it was pertinent that astrophysics teaching and research should be promoted and strengthened in universities. IUCAA was thus conceived as a common university facility. Around the same time, JVN was also thinking to move out of TIFR and was looking for something challenging. On the other hand, Yash Pal was at the command in UGC ever hungry of new ideas and projects. This was a fortuitous circumstance that gave rise to the idea of IUCAA and JVN instantly took up the challenge. The rest followed beautifully.

IUCAA is a fascinating story that would require an article for itself. Even at the risk of being self-congratulatory, I have no hesitation in saying that it is not only a leading institute, but is also a novel experiment in institute governance and participative and democratic functioning. It follows the dictum of '*trust breeds trust*'. All the responsibilities, including financial, are shared by all the faculty members, who actively participate in all decision-making. The Director has very little to do, as I could vouch from my own experience. It is a testimony to the fact that a well-geared responsibility-sharing system has been evolved, such that everything goes on smoothly and effortlessly without anyone getting hassled. It is a visitor's institute and it entertains over thousand visitors a year. Yet it is remarkable how everything is taken care of without any intervention from faculty members. It is to the credit of its support staff that visitors from all over pay very rich and affectionate compliments for hospitality offered to them. The key is to make everyone feel involved in whatever one is doing by giving functional freedom.

Another instance is in order to highlight JVN's commitment to participative and democratic functioning. For faculty hirings, there is a screening committee of senior faculty. If a candidate is found suitable, then referee reports are sought and finally it comes to the selection committee. Once he proposed a name to be considered and sent a mail seeking for a meeting of the screening committee. We all said that the person did not make the grade, he got angry and convened the committee meeting. We all spoke out our views which he patiently listened and then gracefully accepted our view by saying that he just wanted it to be deliberated in the meeting. This is one instance that stands as a sterling example of how true and committed JVN is to what he professes. Would it surprise anyone why he enjoys such trust and unflinching loyalty from his colleagues?

IUCAA is all set for big things and is doing very well. Nothing could be more gratifying than running into a student from Raipur University at one of the largest telescopes in the world in the ATACAMA Chilean desert. That is exactly what JVN had dreamt for IUCAA, that the best astronomical facilities in the world should be accessible to an ordinary university student.

JVN has been a wonderful role model for four generations of young students who have grown up looking up to him. I heard eminent people recalling with great fondness their grandparents blessing them with a wish, 'be like Narlikar'. When he laid down the reins of IUCAA, steering it gloriously for the first 15 years of its existence, JVN has again set out an example, how should one detach from one's own creation with utmost grace and candour. He picked up an office in a quiet corner of the Library and Computer Centre block, and one sees him only at tea time at 10.30 a.m., the time imprinted from the Cambridge days, in the Pendulum court. Else he is completely invisible; again there is something to emulate. It is remarkable that JVN never ceases to be a role model. He is indeed a gentleman scientist and a wise man.

Here is a man who does outstanding research, works with missionary zeal to spread the message of science and its method far and wide, writes books, gives

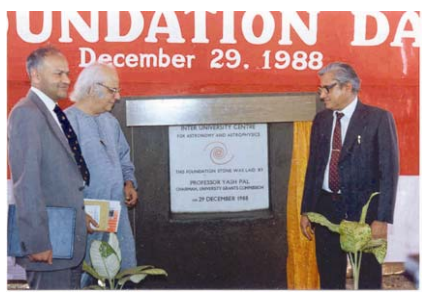

IUCAA Foundation stone laying ceremony.

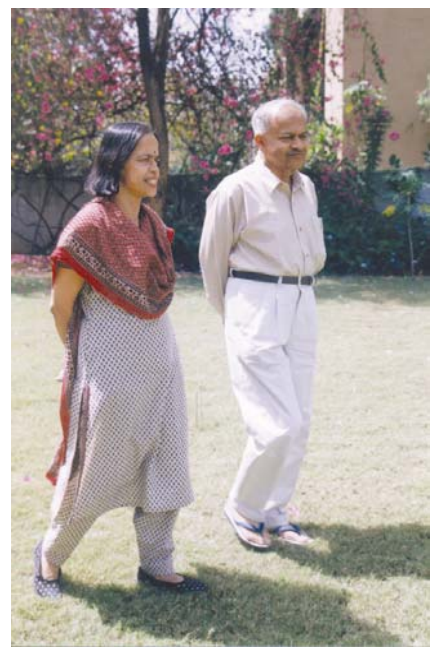

Jayant with his wife.

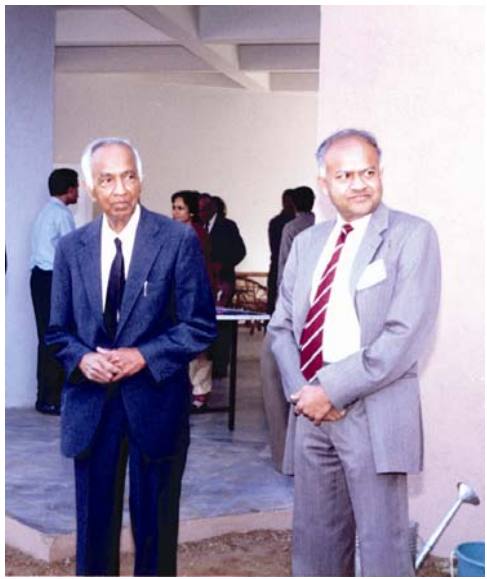

Jayant with S. Chandrasekhar.

lectures and builds up an institute, and also finds time to advise the Government in advisory capacity through various committees. Yet you never find him rushed and hassled. The secret is his peace of mind which is provided in abundance by his wife, Mangala. They are a very fine couple, simple and unassuming and ever ready to help. To put things in right perspective, on one hand, the kind of adulations and admiration he received at a very young age (late twenties; and continues to receive) should have been hard and demanding to digest and carry along all through. And this he has done with remarkable grace and modesty in the best Indian tradition. On the other hand, in his professional work feeling isolated and important contributions not being duly recognized by peers is also equally challenging to cope with. Again he has done it with grace, poise and without rancour. This is the real measure of the man.

On a personal count I should say that it has been the most fascinating and rewarding journey with him and at the slightest hint I would be instantly at his side to team up again. In the ultimate reckoning it does not matter whether one got one's due or not, what matters is how best you lived and worked notwithstanding the rewards and disappointments. Without hesitation I would say that Mangala and JVN have lived outstandingly well and on 19 July 2014, his 76th birthday, I wish him to continue living well.

NARESH DADHICH

*Centre for Theoretical Physics,*
*Jamia Millia Islamia,*
*New Delhi 110 025, India and*
*The Inter-University Centre for*
  *Astronomy and Astrophysics,*
*Post Bag 4, Ganeshkhind,*
*Pune University Campus,*
*Pune 411 007, India*
*e-mail: nkd@iucaa.ernet.in*